\documentclass[pra,twocolumn,superscriptaddress,showpacs]{revtex4}

\usepackage{amsmath,amssymb,graphicx,color,bm,soul}

\begin{document}

\title{Triggered single photon emitters based on stimulated parametric scattering in weakly nonlinear systems}

\author{Oleksandr Kyriienko}
\affiliation{Division of Physics and Applied Physics, Nanyang Technological University 637371, Singapore}
\affiliation{Science Institute, University of Iceland, Dunhagi-3,
IS-107, Reykjavik, Iceland}
\affiliation{QUANTOP, Danish Quantum Optics Center, Niels Bohr Institute, University of Copenhagen, Blegdamsvej 17, DK-2100 Copenhagen, Denmark}

\author{Timothy C. H. Liew}
\affiliation{Division of Physics and Applied Physics, Nanyang Technological University 637371, Singapore}


\begin{abstract}
We introduce a scheme of single photon emission based on four-wave mixing in a three mode system with weak Kerr-type nonlinearity. A highly populated lower energy mode results in strong stimulated scattering of particle pairs out of the central mode, which consequently limits the central mode occupation. Thus, the system can be reduced to a $\chi^{(2)}$ nonlinear medium with greatly enhanced interaction constant. As a model setup we consider dipolaritons in semiconductor microcavities. Using the master equation approach we show strong antibunching under continuous wave pump, which largely exceeds the conventional blockade mechanism. Finally, using a pulsed excitation we demonstrate theoretically a triggered single photon emitter in a weakly nonlinear system with $33\%$ emission probability.
\end{abstract}

\pacs{42.65.Lm,42.50.Dv,71.36.+c}
\maketitle

Single photons~\cite{Shields2007} are an important resource for quantum information technologies~\cite{Scarani2009,Knill2001}, making their production an ultimate challenge of modern quantum optics. Ideally, they can be generated on-demand with high fidelity, device stability, high repetition rate and without post-selection constraints.

Schemes for single photon generation require an optical nonlinearity, which makes the energy spectrum anharmonic in the particle number. This sensitivity can allow the presence of a single photon to block the resonant injection of another~\cite{Imamoglu1997}. However, a necessary condition is that the strength of the nonlinear interaction between two photons exceeds significantly the dissipation rate (linewidth). These values are typically similar in atom-cavity~\cite{Birnbaum2005,Dayan2008} and semiconductor systems~\cite{Reinhard2012}, where the photon blockade was confirmed experimentally. A strong blockade was seen in superconducting circuits, where the nonlinear interaction strength is exceptionally strong~\cite{Lang2011}. Also, the triggered emission of single photons has been achieved in quantum dot~\cite{Michler2000,Santori2001,He2013,Holmes2014,Nowak2014}, atomic~\cite{Kuhn2002McKeever2004} and molecular~\cite{Brunel1999Lounis2000} systems, and p-n junction turnstile devices~\cite{Kim1999}.

Despite these advances, there is still room for improvement in emission quality and efficiency, given the high requirements for quantum applications. A recent theoretical attempt~\cite{Liew2010} at blockade enhancement was based on an unconventional mechanism of quantum interference~\cite{Bamba2011,Carmichael1985} in weakly nonlinear coupled mode systems, where an optical nonlinearity is much smaller than a decay rate of a cavity mode. While originally intended for use in coupled micropillar systems~\cite{Bamba2011b}, similar effects are now being considered in quantum dot cavities~\cite{Majumdar2012}, coupled optomechanical cavities~\cite{Komar2013,Xu2013}, dipolaritons~\cite{Kyriienko2014a}, doubly resonant microcavities~\cite{Gerace2014}, photonic molecules, and passive nonlinear cavities~\cite{Ferretti,Flayac2013,XuLi2014}. A limitation of the scheme is the presence of a fast oscillation in the unequal time second order correlation function, $g_2(\tau)$, which places a strong requirement on the time resolution of detection. Other possible directions to achieve single photon emission in weakly nonlinear systems exploit optomechanical setups at criticality \citep{Lu2014,Xu2014}.

\begin{figure}
\centering
\includegraphics[width=0.75\linewidth]{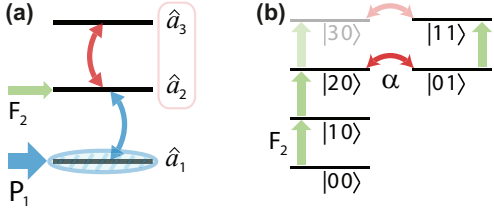}
\caption{(color online) (a) Sketch of the energy levels of the system, showing the parametric scattering between modes. The first mode is strongly driven by the coherent pump $P_1$ and is macroscopically occupied. The second mode is weakly driven by the coherent pump $F_2$. (b) The relevant Fock states corresponding to the $|n_2 n_3\rangle = |n_2 \rangle \otimes |n_3 \rangle$ subspace, where the doubly occupied photon level of the $\hat{a}_2$ mode is coupled to the singly occupied $\hat{a}_3$ mode level, with coupling strength $\alpha$.}
\label{fig:sketch}
\end{figure}

In this work we introduce an alternative route towards the enhancement of single-photon statistics in a weakly nonlinear optical system. We consider three general modes with a Kerr-type nonlinearity and phase matching so as to allow a parametric scattering (four-wave mixing) of particle pairs from the middle mode to upper and lower modes [Fig. \ref{fig:sketch}(a)]. The lower mode of the system is strongly excited, while the middle mode of the system is only weakly excited. The result is a strong limitation of the population of the middle mode, since if there are ever two particles in the middle mode they immediately undergo parametric scattering, which is strongly stimulated by the lower mode amplitude. The system can be envisioned as an effective $\chi^{(2)}$ nonlinear system where conventional parametric photon blockade can be realized~\cite{Majumdar2013}. Here, the condition of strong nonlinearity can be removed by enhancement of the bare interaction due to large occupation of the lower mode.

As a realistic setup for the proposal we consider a system of dipolaritons~\cite{Cristofolini2012} --- mixed modes of strongly coupled cavity photon, direct exciton, and indirect exciton modes. They can be realized in a double quantum well embedded in a semiconductor microcavity~\cite{Cristofolini2012,Christmann2011}, and are useful for various applications~\cite{Kyriienko2013,Kristinsson2013,Kristinsson2014}. The presence of exciton-exciton interactions allows for parametric scattering between dipolariton modes \cite{Kristinsson2014}, and their energy can be easily controlled by an external electric field. Using the parameters of currently existing dipolariton samples, we calculate the second order coherence function for the middle mode, showing that strong antibunching can be realized even for modest values of nonlinearity. Additionally, we demonstrate the triggered emission of single photons under pulsed excitation of the middle mode, which are available with high efficiency ($> 30\%$) and repetition rate (GHz).


Finally, we note that our model has a wide range of other possible physical realizations, including arrangements using coupled photonic crystal fibers~\cite{Langford2011} and cavities~\cite{Azzini2013,Conti2004}, subwavelength grated resonators~\cite{Zhang2013,Kyriienko2014}, and planar semiconductor microcavities, where parametric scattering was observed~\cite{Savvidis2000,Diederichs2005,Ferrier2010,Xie2012,Lecomte2013}.


\textit{Generalized model.} We consider a generic system of three modes coupled by a parametric interaction in the form of four wave mixing. The Hamiltonian of the system reads
\begin{align}
\label{eq:H_start}
\hat{\mathcal{H}} = \sum_{i=1,2,3} E_{i}\hat{a}_i^\dagger \hat{a}_i + \alpha_0 \left( \hat{a}_2^\dagger \hat{a}_2^\dagger \hat{a}_1 \hat{a}_3 + \hat{a}_1^\dagger \hat{a}_3^\dagger \hat{a}_2 \hat{a}_2 \right)\\ \notag + F_2 (\hat{a}_2^\dagger e^{-iE_{F2}t/\hbar} + h.c.) + P_1 (\hat{a}_1^\dagger e^{-iE_{P1}t/\hbar} + h.c.),
\end{align}
where $\hat{a}_i$ ($i=1,2,3$) are annihilation operators of three distinct bosonic modes with energies $E_i$. The second term describes parametric scattering between modes with an interaction constant $\alpha_0$. The last two terms in Eq. (\ref{eq:H_start}) stand for the coherent pumps of the second mode ($F_2$) and first mode ($P_1$) [Fig. \ref{fig:sketch}(a)]. The driving energies are defined as $E_{F2}$ and $E_{P1}$, respectively. The time dependence from Hamiltonian (\ref{eq:H_start}) can be conveniently removed by performing the unitary transformation $\hat{\mathcal{H}}^{\mathrm{(new)}} = \hat{\mathcal{U}}^\dagger (\hat{\mathcal{H}} - i\hbar \partial_t)\hat{\mathcal{U}}$, where the unitary operator reads $\hat{\mathcal{U}}(t) = \exp \{ -i [E_{P1}\hat{a}_1^\dagger \hat{a}_1 + E_{F2}\hat{a}_2^\dagger \hat{a}_2 + (2 E_{F2} - E_{P1}) \hat{a}_3^\dagger \hat{a}_3]t/\hbar \}$. The transformed Hamiltonian yields
\begin{align}
\label{eq:H_trans}
\hat{\mathcal{H}} = \sum_{i=1,2,3} \Delta_{i}\hat{a}_i^\dagger \hat{a}_i + \alpha_0 \left( \hat{a}_2^\dagger \hat{a}_2^\dagger \hat{a}_1 \hat{a}_3 + \hat{a}_1^\dagger \hat{a}_3^\dagger \hat{a}_2 \hat{a}_2 \right)\\ \notag + F_2 (\hat{a}_2^\dagger + \hat{a}_2) + P_1 (\hat{a}_1^\dagger + \hat{a}_1),
\end{align}
where detunings of the modes are $\Delta_1 = E_1 - E_{P1}$, $\Delta_2 = E_2 - E_{F2}$, and $\Delta_3 = E_3 + E_{P1} - 2 E_{F2}$.

We consider the pump of the first mode $P_1$ to be strong, meaning that large occupation of the low energy mode can be achieved. In this case the system can be linearized using the substitution $\hat{a}_i \mapsto \langle \hat{a}_i \rangle + \delta\hat{a}_i$, where $\langle \hat{a}_i \rangle$ denotes the quasiclassical field of mode $i$, and $\delta\hat{a}_i$ denotes the annihilation operator of a quantum fluctuation. We are interested in the situation where the bare parametric scattering constant $\alpha_0$ is much smaller than the pumping rate $P_1$ and the characteristic decay rate of the first mode $\kappa_1$, and the parametric scattering condition is satisfied ($\Delta_{2,3} = 0$). In this case the occupation of the lower mode can be calculated as $n_1 = \langle \hat{a}_1^\dagger \hat{a}_1 \rangle = P_1^2/[\Delta_1^2 + (\kappa_1/2)^2]$. The numerical simulation for dynamical equations for amplitudes $\langle \hat{a}_i \rangle$ shows that with zero pump detuning $\Delta_1$ and $P_1/\kappa_1 = 30$, the lowest mode occupation is $n_1 \approx 3.6 \times 10^3$, while the amplitudes of the second and third modes are negligible, $\langle \hat{a}_{2,3} \rangle \approx 0$. This allows us to approximate the first mode by a quasiclassical amplitude $\langle \hat{a}_1 \rangle = \sqrt{n_1}$ and reduce the consideration to the weakly occupied quantum modes $\hat{a}_{2}$ and $\hat{a}_{3}$. The Hamiltonian (\ref{eq:H_trans}) then becomes:
\begin{align}
\label{eq:H_23}
\hat{\mathcal{H}}' = \sum_{i=2,3} \Delta_{i}\hat{a}_i^\dagger \hat{a}_i + \alpha \left( \hat{a}_2^\dagger \hat{a}_2^\dagger \hat{a}_3 + \hat{a}_3^\dagger \hat{a}_2 \hat{a}_2 \right) + F_2 (\hat{a}_2^\dagger + \hat{a}_2),
\end{align}
where we subtracted the non-relevant overall energy shift coming from macroscopic occupation of the first mode. The parametric scattering (now an effective three wave mixing) between the second and third modes is enhanced by the square root of the occupation of the first mode, $\alpha = \alpha_0 \sqrt{n_1}$. This allows one to manipulate the coupling controlling the pump amplitude $P_1$, and increase the coupling by several orders of magnitude.
\begin{figure}
\includegraphics[width=0.75\linewidth]{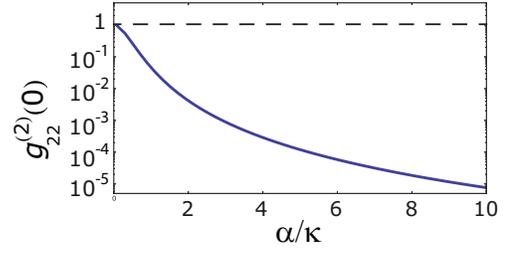}
\caption{(color online) $g^{(2)}_{22}(0)$ plotted as a function of dimensionless parametric scattering rate $\alpha/\kappa$ for $\Delta_2=\Delta_3=0$. Blue solid curve corresponds to cw coherent pump $F_2 /\kappa = 0.1$. Dashed line shows the classical value of $g^{(2)}_{22}(0) = 1$ corresponding to a coherent optical source.}
\label{fig:conv}
\end{figure}

We start by considering the reduced two mode system given by Hamiltonian (\ref{eq:H_23}). This system was studied previously in Ref. \cite{Majumdar2013}, however for a different setup represented by a doubly resonant photonic crystal cavity, where enhancement of the parametric interaction is absent. In order to study the single photon emission from the second mode, we calculate the second order correlation function $g^{(2)}_{22}(\tau)= \langle \hat{a}_2^\dagger(t) \hat{a}_2^\dagger(t+\tau) \hat{a}_2(t+\tau) \hat{a}_2(t) \rangle / ( \langle \hat{a}_2^\dagger(t) \hat{a}_2(t) \rangle \langle \hat{a}_2^\dagger(t+\tau) \hat{a}_2(t+\tau) \rangle )$. This can conveniently be done using the master equation $\dot{\boldsymbol{\rho}}=-i \hbar^{-1} \big[ \hat{\mathcal{H}}', \boldsymbol{\rho} \big] + \hat{\mathcal{L}}\boldsymbol{\rho}$ for the density matrix $\boldsymbol{\rho}$, where $\hat{\mathcal{L}} \boldsymbol{\rho} = \sum_i \kappa_i (\hat{a}_i \boldsymbol{\rho} \hat{a}_i^\dagger - \{ \hat{a}_i^\dagger \hat{a}_i, \boldsymbol{\rho} \}/2)$ corresponds to the Lindblad superoperator, accounting for decay of the modes $\kappa_i$ \cite{Liew2010}.

Considering weak continuous wave (cw) coherent pump $F_2 / \kappa = 0.1$, we calculate the steady state zero delay coherence function $g^{(2)}_{22}(0)$ as a function of enhanced interaction constant $\alpha/\kappa$ ($\kappa_i \equiv \kappa$, $i=1,2,3$). We find that in general non-classical photon statistics can be reached for small detunings, and note that the strongest antibunching can be achieved when $\Delta_{2,3} = 0$. Setting an optimal zero detuning, corresponding to efficient parametric scattering, and considering the weak pump limit $F_2 / \kappa \ll 1$,  the second-order coherence function at zero delay can be derived analytically using the trial wave function approach \cite{Bamba2011,Komar2013}:
\begin{equation}
g^{(2)}_{22}(0)=\frac{1}{1 + 8\alpha^2/\kappa^2 + 16 \alpha^4/\kappa^4}.
\label{eq:g2}
\end{equation}
This closely matches the results of master equation simulation shown in Fig. \ref{fig:conv}, where reduction of the second-order coherence to zero value at large parametric coupling is revealed (blue solid curve). This behavior resembles the conventional photon blockade case, requiring the condition $2 \alpha/\kappa >1$, while keeping the enhancement of initial nonlinearity $\alpha=\alpha_0\sqrt{n_1}$.

The origin of induced single photon emission can be understood from the sketch in Fig. \ref{fig:sketch}(b). The parametric coupling solely acts on the two-photon level of the pumped second mode, transferring particles to the third mode. Namely, the amplitude for the $|20\rangle$ level reads $A_{20} = i \kappa/(2\sqrt{2}\alpha)A_{01}$, showing that two photon occupation $N_{20}=|A_{20}|^2$ can be made sufficiently small comparing to $N_{01}=|A_{10}|^2$, once large coupling $\alpha$ is present.

Next, we proceed with proposing a specific physical setup where the described stimulated single photon emission can be realized.


\textit{Dipolariton system.} We consider a system of coupled quantum wells placed inside a semiconductor microcavity, as considered in Refs. \cite{Cristofolini2012} and \cite{Kristinsson2013}. The Hamiltonian of the system can be divided into linear and nonlinear parts, $\hat{\mathcal{H}}=\hat{\mathcal{H}}_0+\hat{\mathcal{H}}_\mathrm{int}$.
The linear part of the Hamiltonian contains the bare energies of the cavity photon ($E_C$), direct exciton ($E_{DX}$) and indirect exciton ($E_{IX}$) modes, which are represented by field operators $\hat{a}$, $\hat{b}$ and $\hat{c}$, respectively:
\begin{align}
\hat{\mathcal{H}}_0&=E_C\hat{a}^\dagger\hat{a}+E_{DX}\hat{b}^\dagger\hat{b}+E_{IX}\hat{c}^\dagger\hat{c}\notag\\
&+\Omega\left(\hat{a}^\dagger\hat{b}+\hat{b}^\dagger\hat{a}\right)-J\left(\hat{b}^\dagger\hat{c}+\hat{c}^\dagger\hat{b}\right),
\label{eq:H0}
\end{align}
where $\Omega$ and $J$ are the direct exciton-photon light-matter coupling and direct-indirect exciton tunneling coupling, respectively.

The nonlinear part of the Hamiltonian reads
\begin{align}
\hat{\mathcal{H}}_\mathrm{int}&=\alpha_D\hat{b}^\dagger\hat{b}^\dagger\hat{b}\hat{b}+\alpha_I\hat{c}^\dagger\hat{c}^\dagger\hat{c}\hat{c}+\alpha_{DI}\hat{b}^\dagger\hat{c}^\dagger\hat{b}\hat{c},
\label{eq:Hint}
\end{align}
where $\alpha_D$, $\alpha_I$ and $\alpha_{DI}$, represent the strength of interactions between pairs of direct excitons, pairs of indirect excitons, and direct-indirect exciton pairs, respectively.
The evolution of the system is given by the master equation for the density matrix:
\begin{align}
&i\hbar\frac{d\boldsymbol{\rho}}{dt}=\left[\mathcal{H},\boldsymbol{\rho}\right]+\frac{i\Gamma_C}{2}\left(2\hat{a}\boldsymbol{\rho}\hat{a}^\dagger-\hat{a}^\dagger\hat{a}\boldsymbol{\rho}-\boldsymbol{\rho}\hat{a}^\dagger\hat{a}\right)\\ \notag
&+\frac{i\Gamma_X}{2}\left(2\hat{b}\boldsymbol{\rho}\hat{b}^\dagger+2\hat{c}\boldsymbol{\rho}\hat{c}^\dagger-(\hat{b}^\dagger\hat{b}+\hat{c}^\dagger\hat{c})\boldsymbol{\rho}-\boldsymbol{\rho}(\hat{b}^\dagger\hat{b}+\hat{c}^\dagger\hat{c})\right),
\end{align}
where dissipation has been introduced in the Lindblad form where $\Gamma_C$ and $\Gamma_X$ are the decay rates of cavity photons and excitons respectively (for simplicity, we assume that direct and indirect excitons have the same decay rate).

The linear part of the Hamiltonian, $\mathcal{H}_0$, can be diagonalized by introducing the dipolariton field operators, $\hat{A}_i$:
\begin{align}
\hat{a}&=V_{11}\hat{A}_1+V_{21}\hat{A}_2+V_{31}\hat{A}_3, \label{eq:a}\\
\hat{b}&=V_{12}\hat{A}_1+V_{22}\hat{A}_2+V_{32}\hat{A}_3, \label{eq:b}\\
\hat{c}&=V_{13}\hat{A}_1+V_{23}\hat{A}_2+V_{33}\hat{A}_3, \label{eq:c}
\end{align}
where $V_{ij}$ are the matrix elements of eigenvectors, corresponding to dipolariton Hopfield coefficients \cite{Byrnes2014}.
\begin{figure}
\includegraphics[width=1.0\linewidth]{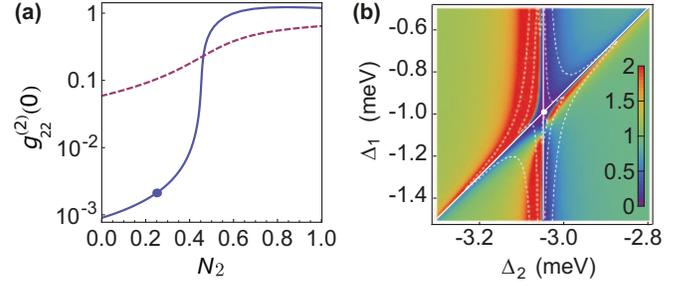}
\caption{(color online) (a) Dependence of $g^{(2)}_{22}(0)$ on the occupation $N_2$ for the three mode (solid) and single mode blockade (dashed). (b) Dependence of $g^{(2)}_{22}(0)$ on the pump detunings $\Delta_1 =E_1 - E_{P1}$ and $\Delta_2 =E_2 - E_{F2}$, plotted for low pump intensity $F_2$. The solid lines show the conditions $\Delta_2+c_1|\psi_1|^2 = 0$ and $\Delta_3 + c_2|\psi_1|^2= 0$, where the optimum antibunching occurs at their intersection. The dashed curves show contours in the average occupation of the upper mode.}
\label{fig:g2DensityMatrix}
\end{figure}

Using this substitution and Eqs.~(\ref{eq:a})--(\ref{eq:c}), the Hamiltonian written in the basis of lower ($\hat{A}_1$), middle ($\hat{A}_2$) and upper ($\hat{A}_3$) dipolaritons becomes:
\begin{align}
\notag
\mathcal{H}_{\mathrm{dpl}}&=\sum_{j=1,2,3}E_j\hat{A}^\dagger_i\hat{A}_i+\left(c_1\hat{A}_2^\dagger\hat{A}_2\hat{A}_1^\dagger \hat{A}_1 + c_2\hat{A}_3^\dagger\hat{A}_3 \hat{A}_1^\dagger \hat{A}_1 \right)\\ \notag
&+c_3\hat{A}_2^\dagger\hat{A}_2^\dagger\hat{A}_2\hat{A}_2+c_4\hat{A}_3^\dagger\hat{A}_3^\dagger\hat{A}_3\hat{A}_3+c_5\hat{A}_2^\dagger\hat{A}_3^\dagger\hat{A}_2\hat{A}_3\\
\label{eq:Dipolariton}
&+c_6\left(\hat{A}_2^\dagger\hat{A}_2^\dagger \hat{A}_3\hat{A}_1 + \hat{A}_3^\dagger \hat{A}_1^\dagger \hat{A}_2\hat{A}_2 \right) \\ \notag
&+ P_1 (\hat{A}_1^\dagger e^{-iE_{P1}t/\hbar} + h.c.) + F_2 (\hat{A}_2^\dagger e^{-iE_{F2}t/\hbar} + h.c.),
\end{align}
where $E_j$ are the eigenenergies of $\mathcal{H}_0$, and $c_j$ are given by:
\begin{align}
c_1=&4\alpha_DV^2_{12}V^2_{22}+4\alpha_IV^2_{13}V^2_{23}+\alpha_{DI}\left(V_{12}V_{23}+V_{22}V_{13}\right)^2,\notag\\
c_2=&4\alpha_DV^2_{12}V^2_{32}+4\alpha_IV^2_{13}V^2_{33}+\alpha_{DI}\left(V_{12}V_{33}+V_{32}V_{13}\right)^2,\notag\\
c_3=&4\alpha_DV^4_{22}+\alpha_IV^4_{23}+\alpha_{DI}V^2_{22}V^2_{23}\notag,\\
c_4=&4\alpha_DV^4_{32}+\alpha_IV^4_{33}+\alpha_{DI}V^2_{32}V^2_{33}\notag,\\
c_5=&4\alpha_DV^2_{22}V^2_{32}+4\alpha_IV^2_{23}V^2_{33}+\alpha_{DI}\left(V_{22}V_{33}+V_{32}V_{23}\right)^2\notag,\\
c_6=&2\alpha_DV^2_{22}V_{12}V_{32}+2\alpha_IV^2_{23}V_{13}V_{33},\notag\\
&+\alpha_{DI}V_{22}V_{23}\left(V_{12}V_{33}+V_{32}V_{13}\right).
\label{eq:c1-6}
\end{align}
The last line of Hamiltonian (\ref{eq:Dipolariton}) corresponds to the strong drive of lower dipolariton mode with intensity $|P_1|^2$ and energy $E_{P1}$, and the weak coherent pumping of middle dipolariton mode of $|F_2|^2$ intensity at energy $E_{F2}$. It is assumed that $E_{F2}$ is tuned close to $E_2$ and far from $E_1$ and $E_3$ so as to cause a direct excitation of the middle branch only.

Performing the unitary rotation in order to eliminate the time dependence of the Hamiltonian (\ref{eq:Dipolariton}), similarly to the generic model introduced in the previous section, and assuming the strong pump amplitude for the lowest dipolariton mode ($P_1$ is much greater than the characteristic decay rate of the modes), we can write the effective dipolariton Hamiltonian:
\begin{align}
\label{eq:DipolaritonHeff}
\mathcal{H}_{\mathrm{eff}}&=\sum_{j=2,3}\Delta_{j}\hat{A}^\dagger_i\hat{A}_i+\left(c_1\hat{A}_2^\dagger\hat{A}_2+c_2\hat{A}_3^\dagger\hat{A}_3\right)|\psi_1|^2\\ \notag
&+c_3\hat{A}_2^\dagger\hat{A}_2^\dagger\hat{A}_2\hat{A}_2+c_4\hat{A}_3^\dagger\hat{A}_3^\dagger\hat{A}_3\hat{A}_3+c_5\hat{A}_2^\dagger\hat{A}_3^\dagger\hat{A}_2\hat{A}_3\\ \notag
&+c_6\left(\hat{A}_2^\dagger\hat{A}_2^\dagger\hat{A}_3\psi_1 +\hat{A}_2\hat{A}_2\hat{A}_3^\dagger \psi_1^*\right),
\end{align}
where $\psi_1$ denotes the classical amplitude for the macroscopically occupied lower dipolariton mode, and $|\psi_1|^2 = N_1$ corresponds to its occupation number. The detunings are $\Delta_2 = E_2 - E_{F2}$, $\Delta_3 = E_3 + E_{P1} - 2 E_{F2}$, and $\Delta_1 = E_1 - E_{P1}$. We note that the last term in Eq.~(\ref{eq:DipolaritonHeff}) is of the same form as the generic parametric interaction term considered in Eq.~(\ref{eq:H_23}).

The master equation becomes
\begin{align}
\label{eq:master_eff}
i\hbar\frac{d\boldsymbol{\rho}}{dt}=&\left[\mathcal{H_\mathrm{eff}},\boldsymbol{\rho}\right]\\ \notag
&+\sum_{j=2,3}\frac{i\Gamma_j}{2}\left(2\hat{A}_j\boldsymbol{\rho}\hat{A}^\dagger_j-\hat{A}_j^\dagger\hat{A}_j\boldsymbol{\rho}-\boldsymbol{\rho}\hat{A}_j^\dagger\hat{A}_j\right),
\end{align}
where $\boldsymbol{\rho}$ now represents the density matrix on the reduced subspace spanned by modes $\hat{A}_2$ and $\hat{A}_3$. The decay rates of the modes are given by $\Gamma_j=V^2_{j1}\Gamma_C+\left(V^2_{j2}+V^2_{j3}\right)\Gamma_X$.
%
\begin{figure}
\includegraphics[width=0.75\linewidth]{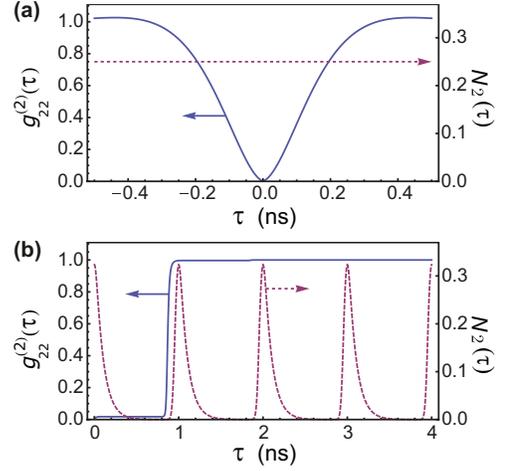}
\caption{(color online) Time delay dependence of $g^{(2)}_{22}(\tau)$ (solid traces) and $N_2(\tau)$ (dashed traces) under continuous wave (a) and pulsed (b) excitation. The pulse duration was set to $50$ ps and the pulse repetition rate to $1~\mu$s.}
\label{fig:g2t}
\end{figure}

We consider the realistic dipolariton system with the following parameters \cite{Cristofolini2012}: $\hbar/\Gamma_C=2.5$~ps, $\hbar/\Gamma_X=500$~ps, $\alpha_D=0.004$~meV, $\alpha_I=0.016$~meV, $\alpha_{DI}=0.008$~meV, $\psi_1=50$, $\Omega=6$~meV, $J=3$~meV, $E_{DX}-E_{IX}=9$~meV, $E_C-E_{IX}=-9$~meV, with a micropillar sample diameter being $2~\mu$m. The last parameters, being detunings between modes, can be controlled by the applied electric field which modifies the energy of the indirect exciton $E_{IX}$. From this we can deduce the effective parameters $c_j$ and $\Gamma_j$, which are given by $c_3=0.0143$ meV, $c_4=0.0035$ meV, $c_5=-0.0027$ meV, $c_6|\psi_1|^2=-1.7256$ meV, $\Gamma_2=0.0072$ meV and $\Gamma_3=0.0207$ meV.

The variation of the second order correlation function of the middle polariton mode with zero time delay $g^{(2)}_{22}(0)$ on the mode occupation $\langle \hat{A}_2^\dagger \hat{A}_2 \rangle \equiv N_2$ is shown in Fig.~\ref{fig:g2DensityMatrix}(a). It reveals the strong dependence of the coherence properties on the pump strength $F_2/\kappa$, showing that high fidelity of the single photon emission source can be achieved. For the considered lower mode occupation values of $g^{(2)}_{22}(0)$ are several orders lower than in the case of the single mode conventional blockade (where $\psi_1=0$ and $\Delta_2 = 0$).

The variation of $g^{(2)}_{22}(0)$ with the pump detunings $\Delta_1 = E_1 - E_{P1}$ and $\Delta_2=E_2 - E_{F2}$ is shown in Fig.~\ref{fig:g2DensityMatrix}(b). The optimum point can be reached by varying the two laser detunings. The white dashed contours show the occupation $N_3$, indicating that the antibunching in the middle polariton mode is associated with parametric scattering to the upper (and lower) polariton modes.

The second order correlation function with non-zero time-delay is shown in Fig.~\ref{fig:g2t} under both continuous wave [Fig.~\ref{fig:g2t}(a)] and pulsed [Fig.~\ref{fig:g2t}(b)] excitation of the middle mode. In contrast to mechanisms of antibunching based on quantum interference, there are no fast oscillations in $g^{(2)}_{22}(\tau)$ under continuous wave pumping and time resolution on the order of hundreds of picoseconds would be sufficient for detection of the effect. Under pulsed operation, $g^{(2)}_{22}(\tau)$ remains close to zero throughout the duration of each pulse, indicating that no more than one photon is emitted per pulse. The short lifetime of the system allows the time of single photon emission to be determined to an accuracy of $\sim 200$ ps and the device can be operated with repetition rates on the order of gigahertz. In the same time we underline that the demonstrated single photon emission source is a probabilistic one, with fidelity of SPE given by the occupation probability of the antibunched mode (around $33\%$). Moreover, analyzing the data from Fig. \ref{fig:g2DensityMatrix}(a) we find that the emission probability of the source can be increased to $45\%$, while keeping reasonably low second-order coherence $g^{(2)} < 0.1$.

Finally, we note that while spontaneous emission of photons occurs from all dipolariton modes, we are only interested in the antibunched output of the middle mode. Therefore, a spectral filtering shall be used in order to observe the devised single photon emission.

To conclude, we have proposed a scheme of single photon emission based on three parametrically coupled modes. Under strong pump of the lowest energy mode, we showed that a weakly pumped middle mode can emit anti-bunched light even for a small value of bare nonlinearity. This can be explained reducing the system to an effective $\chi^{(2)}$ nonlinear setup with enhanced parametric interaction. Considering the system of dipolaritons, we have shown that strong antibunching can be observed for the realistic setup with a weak nonlinearity. Finally, we demonstrated that controlling the intensity of a weak pump a triggered single photon emitter can be realized.

We thank I. A. Shelykh for useful discussions. O.K. acknowledges support from FP7 ITN NOTEDEV network, FP7 IRSES project POLATER, and Eimskip fund. T.L. acknowledges support from the Lee Kuan Yew Endowment Fund.

\end{document}